\documentclass[fleqn,usenatbib]{mnras}

\usepackage{newtxtext,newtxmath}

\usepackage[T1]{fontenc}
\usepackage{ae,aecompl}
\usepackage{hyperref}

\usepackage{graphicx}	
\usepackage{amsmath}	
\usepackage{amssymb}	

\title[$\gamma^2$ Vel region revealed by Gaia]{A sextet of clusters in the Vela OB2 region revealed by Gaia\thanks{Based on observations collected at the European Organisation for Astronomical Research in the Southern Hemisphere under ESO programme(s)
188.B-3002, 096.C-0730(A) and 097.C-0749(A).}}
\author[G. Beccari et al.]{Giacomo Beccari,$^{1}$\thanks{E-mail: gbeccari@eso.org}
Henri M.J. Boffin,$^{1}$
Tereza Jerabkova,$^{1,2,3}$
Nicholas J. Wright$^{4}$,
\newauthor
Venu M. Kalari$^{5}$,
Giovanni Carraro$^{6}$,
Guido De Marchi$^{7}$ and
Willem-Jan de Wit$^{8}$
\\
$^{1}$European Southern Observatory, Karl-Schwarzschild-Strasse 2, 85748 Garching bei M\"unchen\\
$^{2}$Helmholtz Institut f\"{u}r Strahlen und Kernphysik, Universit\"{a}t Bonn, Nussallee 14--16, 53115 Bonn, Germany\\
$^{3}$Astronomical Institute, Charles University in Prague, V Hole\v{s}ovi\v{c}k\'ach 2, CZ-180 00 Praha 8, Czech Republic\\
$^{4}$Astrophysics Group, Keele University, Keele, ST5 5BG, UK\\
$^{5}$Departamento de Astronomia, Universidad de Chile, Casilla 36-D, Correo Central, Santiago, Chile\\
$^{6}$Dipartimento di Fisica e Astronomia Galileo Galilei, Vicolo Osservatorio 3, I-35122, Padova, Italy\\
$^{7}$Research \& Scientific Support Department, ESA ESTEC, Keplerlaan 1, 2200 AG Noordwijk, The Netherlands\\
$^{8}$European Southern Observatory, Alonso de Córdova 3107, Casilla 19001, Santiago, Chile
}

\date{Accepted XXX. Received YYY; in original form ZZZ}

\pubyear{2018}

\begin{document}
\label{firstpage}
\pagerange{\pageref{firstpage}--\pageref{lastpage}}
\maketitle

\begin{abstract}
{Using Gaia DR2 data, combined with OmegaCAM ground-based optical photometry from the AD-HOC survey,
and detailed Radial Velocity measurements from ESO-Gaia, we analyse in detail a 10$\times$5$~\deg$ 
region around the Wolf-Rayet star $\gamma^2$ Vel, including the previously known clusters Gamma Vel and NGC2547. 
Using clustering analysis that considers positions, proper motions and parallax, we discover 6 clusters or associations -- 4 of 
which appear new. Analysis of the colour-magnitude diagram for these clusters show that 4 of them formed coevally from the same molecular clouds 10 Myr ago, 
while NGC 2547 formed together with a newly discovered cluster 30 Myr ago. This study shows the incredible wealth of data provided by Gaia 
for the study of young stellar clusters.}
\end{abstract}

\begin{keywords}
{Stars: formation -- Stars: pre-main sequence -- Open clusters and associations}
\end{keywords}



\section{Introduction}

Young clusters are a formidable tool to study star formation, stellar evolution, binary stars, as well as the formation and evolution of clusters themselves. 
They thus offer a window to observationally answer questions in many astrophysical domains. Hence, it is 
critical to discover and study as wide a range of clusters as possible. 

We recently concluded the Accretion Disk with OmegaCAM (ADHOC) survey on the VLT Survey 
Telescope (VST), whose main goal is to study the population of Pre-Main sequence
stars identified using H$\alpha$ excess emission as signature of on-going accretion. As part of the survey, we observed
a 10$\times$5$~\deg$ region around the Wolf-Rayet Star $\gamma^2$ Vel. 
This region has received considerable attention in the recent years after the study by~\citet[][]{jeff14} which, using Radial Velocity (RV) measurements
from the Gaia-ESO survey~\citep[GES][]{gil12,ran13} found that the cluster around $\gamma^2$ Vel (namely Gamma Vel) is in fact composed of two 
coeval but kinematically distinct populations A and B. 
Interestingly~\citet[][]{sacco15} used GES observations to study the stellar population on a region centered  on NGC 2547, a $\approx$35 Myr cluster located two degrees 
south of $\gamma^2$ Vel. They find a population of stars whose RVs, lithium absorption lines, and positions in a colour-magnitude diagram (CMD)
are consistent with those of Gamma Vel B, which they call NGC 2547 B. 
It is hence likely that the stars observed by~\citet[][Gamma Vel B]{jeff14} and~\citet[][NGC 2547 B]{sacco15} in fact belong to a young, low-mass stellar population spread over at least several square degrees in the Vela OB2 complex. Such result has been supported via a set of N-BODY simulation from~\citet[][]{map15}
and might imply that stars in NGC 2547 B could have originally been part of a cluster around $\gamma^2$ Velorum that expanded after gas expulsion or formed in a less dense environment that is spread over the whole Vela OB2 region~\citep[][]{pri16}.
More recently~\citet[][]{dam17} used the GAIA DR1 catalogue TGAS to perform a proper-motion (PM) study of the stellar population in a radius of 4 degrees from the midpoint between the $\gamma^2$ Vel and the NGC 2547 center. They find 2 clear and well distinguished stellar populations in
the PM space offered by TGAS namely C and D. Using GAIA-G  and 2MASS Ks band photometry, they suggest that 
population-D is older than C. Hence, population D is likely consistent with NGC 2547 while population C seems to belong to the Gamma Velorum cluster. Still, no firm conclusion could be drawn based on the GAIA DR1 data. \citet[][]{arm18} traced the spatial distribution of low-mass stars across Vela OB2 using Gaia DR1 photometry and identified a number of spatial overdensities hinting at an extended low-mass population in the area.\\
 The recent second Gaia data release \citep[DR2][]{GAIA_DR2} offers an unique opportunity to characterise and discover new stellar clusters thank to the very precise parallaxes and PMs. Here, we use Gaia DR2 data, combined with new ground-based optical photometry to study a wide region around $\gamma^2$ Velorum and determine precise properties of several clusters or associations therein.

\section{Photometric and astrometric data} \label{Sec:Data}


\label{sec:data_sample}
We secured with OmegaCAM, attached to the 2.4-m VST telescope in Paranal, a  set of deep multi-exposures  of a region covering $126  < \alpha < 116 \, \mathrm{deg.}$ and $-50 < \delta <  -44.5 \, \mathrm{deg}$. 
Data were obtained with the $u$, $g$, $r$, $i$ and $H\alpha$ filters, although we will here use mostly the images obtained with the $r$- and $i$ filters. 
We acquired $2\times25$s exposures for each pointing in the $r$ and $i$ band.
Data reduction was carried out at the Cambridge Astronomical Survey Unit (CASU) and we downloaded 
the astrometrically and photometrically calibrated single band 
catalogs from the VST archive at CASU\footnote{\url{http://casu.ast.cam.ac.uk/vstsp/}}. We used a large number of 
stars in common with the AAVSO Photometric All-Sky Survey (APASS) to correct for possible residual offsets in 
$r$ and $i$ magnitudes between each single exposures. The final catalogue includes 1,847,804 objects homogeneously sampled 
in $r$ and $i$ bands down to $r \approx 21.5\, \mathrm{mag}$ in ABMAG.

We retrieved from the Gaia science archive\footnote{\url{http://gea.esac.esa.int/archive/}} all the objects detected by Gaia that are within 10 degrees on the sky from  $\alpha$$ \approx 120.4 \, \mathrm{deg.}$, $\delta$ $ \approx -47.5 \, \mathrm{deg.}$ without any additional filtering. 
We then used the $C^3$ python script from \citep{Riccio2017} in order to identify the stars in common between the OmegaCAM and the Gaia catalogue. We found 1,785,477 targets in common between Gaia and OmegaCAM. We note here that all stars
with magnitude $r>20$ are lost in this process because of the detection limit of the GAIA DR2 catalogue. Finally the GES provides us with RVs for 1369 targets in our catalogue.

\begin{figure*}
\centering
\includegraphics[width=1\hsize]{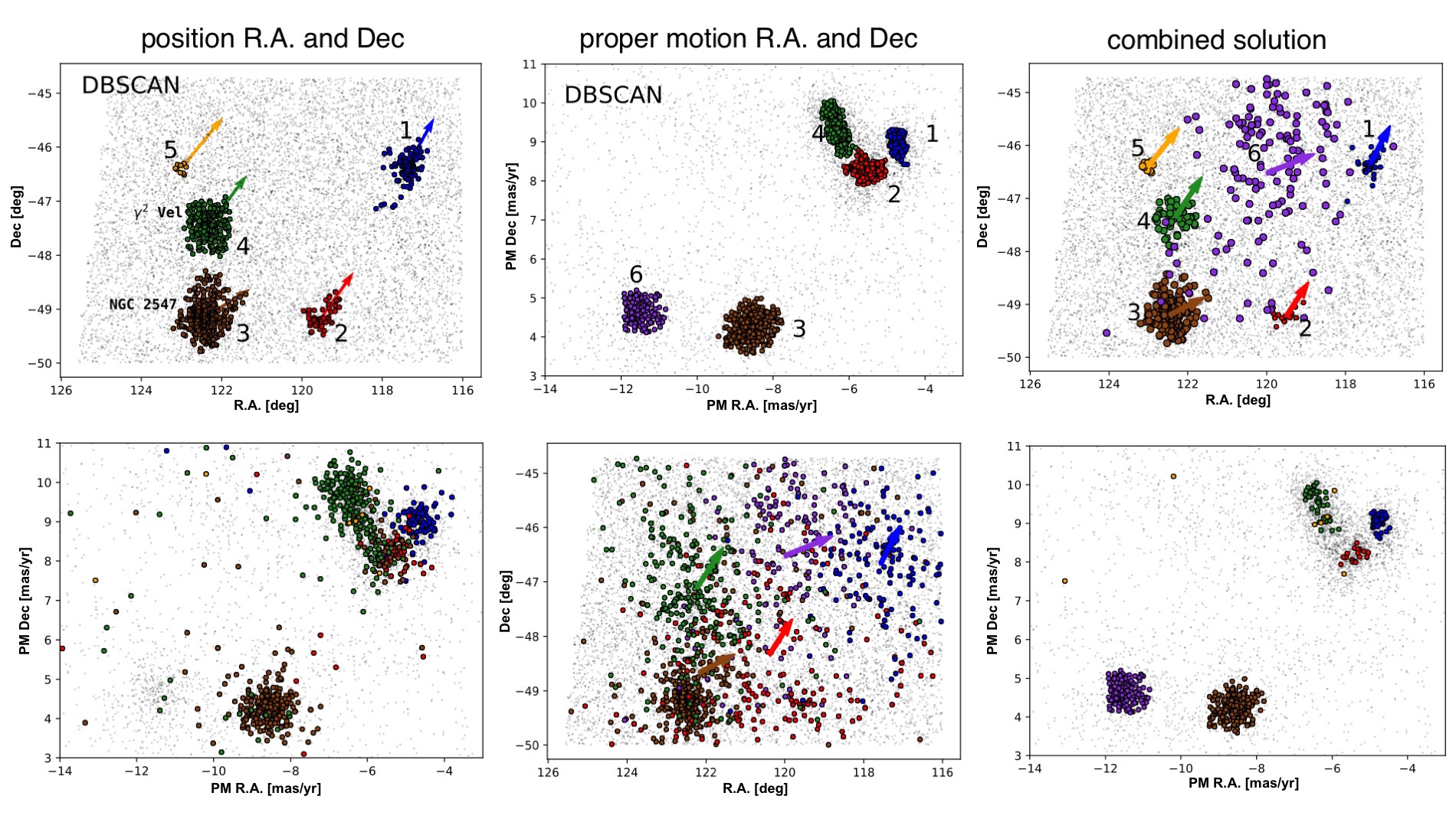}
  \caption{\textit{Left-column}: (top) Cluster selection in the ($\alpha$=R.A., $\delta$=Dec) space with the arrow indicating the mean motion in the plane of the sky and centred on the mean position of the cluster. The same stars are also shown in the proper motion space (bottom-left). Cluster 1 is shown in blue, 2 in red, 3 in brown, 4 in green and 5 in orange. \textit{Central-column}:  (top) Cluster selection in the ($\mu_\alpha$=PM R.A, $\mu_\delta$=PM Dec) space. The new cluster 6 is shown in violet. The same stars are shown in the $\alpha$/$\delta$ plan (bottom). \textit{Central-column}:  Location in  $\alpha$ and $\delta$ (top) and in PM R.A and PM Dec (bottom) of the stars which are in common between the clustering selections applied in the previous plots.}
  \label{fig:pm}
\end{figure*}

\section{Finding clusters}

The data-set that we have collected so far is ideal to investigate the stellar population in the Vela OB2 region. On one side, the Gaia DR2 in fact provides for all the stars in the sampled region, 5 parameter-solutions $\alpha$, $\delta$, PMs in $\alpha$ and $\delta$ ($\mu_\alpha$ and $\mu_\delta$), as well as the parallax, $\varpi$. On the other hand, the OmegaCAM data-set provides accurate multi-band photometric measurements. These data can hence be used to identify sub-populations via clustering algorithms and to study their properties (e.g. age) through the CMD. We want to stress here that in this work we are mainly interested in finding groups of stars whose astrometric and photometric properties allow us to infer their membership to a specific sub-population or association in the studied field, and we are not aiming at completeness. In this respect, we will use the term cluster to indicate a population of stars showing common astrometric and photometric properties. We will study the dynamical state of the sub-populations to assess their dynamical state (e.g. virial status) in a follow-up work.

\subsection{In position and parallax}

To find clusters, we use the python implementation in sci-kit learn\footnote{\url{http://scikit-learn.org/stable/modules/generated/sklearn.cluster.DBSCAN.html}} \citep{scikit-learn},
of the DBSCAN  data clustering algorithm \citep{Ester96adensity-based}. 
As a first step, we select only those stars that have a relative error on the parallax smaller than 10\%, i.e. $\sigma_\varpi / \varpi < 0.1$. Since we are mostly interested in clusters located around the star forming region surrounding $\gamma^2$~Vel, we made a cut in parallax, $2.2 < \varpi < 3.9$mas/y, as $\gamma^2$~Vel has a parallax of 2.92$\pm$0.30~mas/y \citep{Hipparcos07}. The stars in the final catalogue show an average value of $visibility\_periods\_used=15$. Using equation (1)\footnote{Equation (1) presents $u$-parameter that, combining the $astrometric\_chi2\_al$ and $astrometric\_n\_good\_obs\_al$ parameters, ensures that we select only the targets with good astrometric solution.} from \citet[][]{Arenou2018} and filtering for duplicate targets, we remove only 3\% of stars. We do not consider any photometric filter introduced by \citet[][]{Arenou2018} as we use the OmegaCAM photometry.

First, we were interested in looking for clusters in the sky position, and we thus applied DBSCAN in the 3D space, ($\alpha$, $\delta$, $\varpi$), rescaling these three sets of data such that they all had a mean of zero and a standard deviation of 1, and used a scale of 0.25 and a minimum sample of 20. We note here that we have checked that we obtain similar results when using a different value for the minimum sample, and adjust accordingly the scale factor. The clustering method allowed us to distinguish 5 sub-populations on sky (see top left panel of Fig.~\ref{fig:pm}), one of which (Cl5) is however very poorly populated. We show in the bottom panel on the left-hand of Fig.~\ref{fig:pm}, the PMs of the stars belonging to the five clusters discovered in $\alpha$ and $\delta$. Clearly, four of the sub-groups also appear clustered in PM space and are thus indeed kinematically compact. 
The arrows shown in some of the panels indicate the mean motion in the plane of the sky and are centred on the mean position of the sub-group. Quite amazingly, three of the clusters (Cl 1, 2 and 4) have very similar PMs. The smallest concentration we found (Cl 5) contains only 9 stars but several of them have also similar PMs as Cl 4. Clusters 1, 2, 4, and 5 are also evidence on the density map of low-mass stars in Vela OB2 identified by~\citet[][]{arm18}. 
 
\begin{table*}
\caption{Mean properties of the six clusters.}
\label{tab:properties}
\centering 
\begin{tabular}{l c c c c c c c c c}
\hline\hline 
Cl. ID  & $\alpha$&		$\delta$		&	     PM $\alpha$          &        PM $\delta$          &       Parallax        &       Distance            &     Rad. Vel.     & Age & Other Name     \\
 & (deg.) & (deg.) & (mas/yr) & (mas/yr) & (mas) & (pc) & (km/s) & Myr & \\
\hline                                  
1 &      117.366448 & $      -46.339979 $ & $ -4.76 \pm  0.12 $& $ 9.05 \pm  0.17 $&$  2.52 \pm  0.04 $&$   397 $& -- &    10 & \\
2 &      119.566307 & $      -49.212074 $ & $ -5.41 \pm  0.19 $& $ 8.18 \pm  0.15 $&$  2.42 \pm  0.05 $&$   413 $&$  20.36 \pm 0.06 $ &    10 & \\
3 &      122.468165 & $      -49.176393 $ & $ -8.57 \pm  0.29 $& $ 4.27 \pm  0.27 $&$  2.54 \pm  0.07 $&$   393 $&$  12.29 \pm   3.71 $ &    30 & NGC 2547\\
4 &      122.343657 & $      -47.350962 $ & $ -6.40 \pm  0.20 $& $ 9.56 \pm  0.32 $&$  2.85 \pm  0.08 $&$   351 $&$  18.50 \pm   4.50 $ &    10 & Gamma Vel\\
5 &      123.013175 & $      -46.400231 $ & $ -7.52 \pm  2.36 $& $ 9.22 \pm  1.17 $&$  2.74 \pm  0.05 $&$   365 $&    -- &    10 & \\
6 &      119.999214 & $      -46.525013 $ & $ -11.47 \pm  0.26 $& $ 4.65 \pm  0.27 $&$  3.02 \pm  0.07 $&$   331 $&$  11.41 \pm   4.20 $ &    30 & \\
\hline  
\end{tabular}
\end{table*}

\begin{figure}
\centering
\includegraphics[width=0.9\hsize]{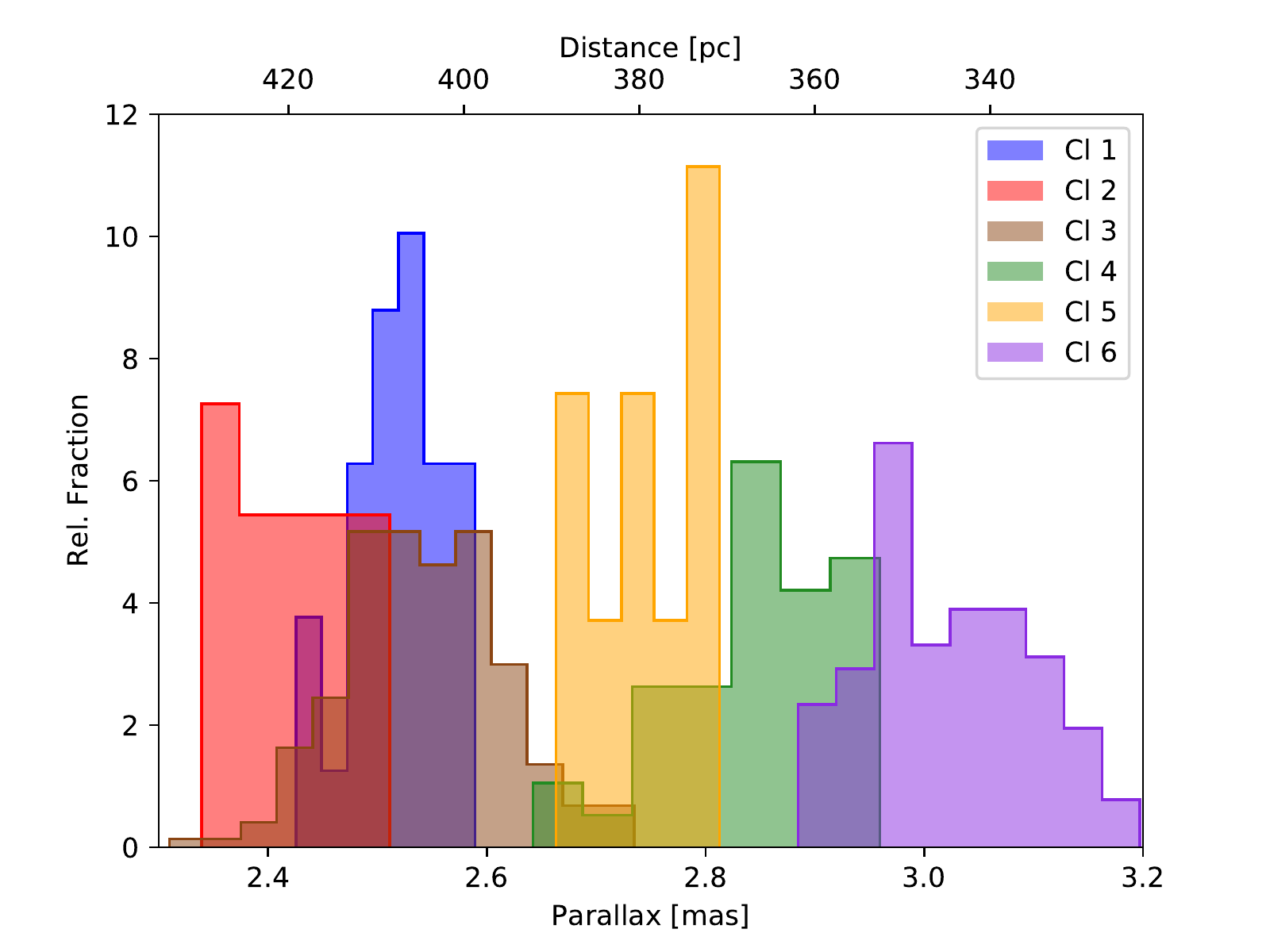}\\  \includegraphics[width=0.9\hsize]{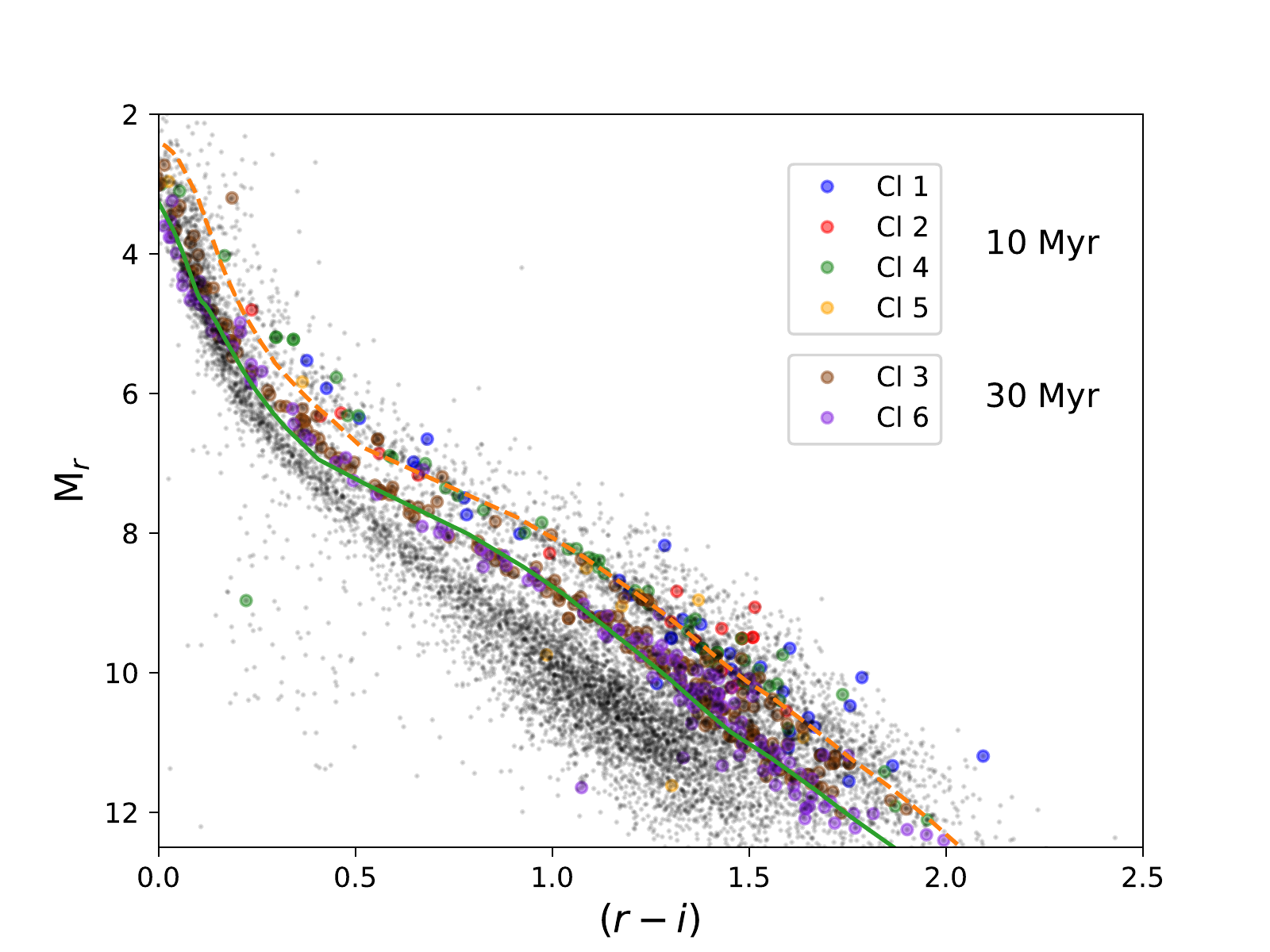}
  \caption{(top) The parallax distribution of the 6 clusters discussed in this paper. The distributions have been normalised. (bottom) The colour-magnitude $(r-i)-M_r$ diagram for all stars in the field with $2.2 < \varpi < 3.9$ and $\sigma_\varpi/\varpi < 0.1$ (black dots) and the stars of the clusters. 
In all plots, the colours correspond to those shown in Fig.~\ref{fig:pm}. 
 }
  \label{fig:parcmdha}
\end{figure}

\subsection{In proper motions}

The inspection of the bottom panel on the left-hand of Fig.~\ref{fig:pm} clearly reveals that there is a cluster in PM space that was not caught via the selection in $\alpha$ and $\delta$ (see small group in the lower left). We thus also looked at clustering in this space, by running DBSCAN in the ($\mu_\alpha$, $\mu_\delta$, $\varpi$) space. Here also five clusters are clearly identified (see top panel in the central column of Fig.~\ref{fig:pm}), four of which are in common with the ones found with the previous method. Interestingly, while Cl 5 is not found in PM, as expected, a new cluster (Cl 6) was found at $\mu_\alpha\approx -11.5$mas/y, $\mu_\delta\approx 4.6$mas/y, that corresponds to stars more loosely spread on the sky. We show in the lower panel of the central column of the figure, the position in $\alpha$ and $\delta$ of the stars selected in PM space. The fact that some clusters are found in PM space but not in $\alpha$/$\delta$ space (and vice versa)
clearly demonstrates that we should not be looking at clusters in 5-dimensional space (positions, PMs, parallaxes) only, but that the spatial
information should be consistently checked.

As final step in our study, we have then taken a conservative approach and retained, for the 4 clusters where this applies (namely Cl 1,2,3 and 4), only those stars that are selected using both methods. This allows us to have a much better indication on the location and motion of the clusters, at the expense of its completeness. We show in the right-hand column of Fig.~\ref{fig:pm} the position of the final selection of sub-groups in the $\alpha$/$\delta$ (top right plot) and PM space (bottom right plot). We list in Table~\ref{tab:properties} the mean properties for the final set of clusters identified with the combination of the two approaches.

\section{Properties of the clusters}

We will discuss from now on, only the stars belonging to the sub-populations (or clusters) as shown in the plots in the right-hand column of Fig.~\ref{fig:pm}. 
It is immediately obvious that Cl 3 and 4 coincide with the well known clusters NGC 2547 and Gamma Vel while the position of the other clusters does not correspond to any known cluster.
In Fig.~\ref{fig:parcmdha}, we show the parallax distribution and the CMD of the 6 clusters. Given the fact that the clustering was made in a 3D space that included the parallax, it is no surprise that each cluster corresponds to a well defined (and limited) parallax range. Given that we only use those stars which have precise parallaxes, we can simply infer the distances from these, without needing any prior knowledge on the Bayesian probabilities of our targets. It is seen that 3 clusters (Cl 1, 2, 3) are at similar distances ($\approx 400$~pc), while the 3 others are closer to us (between 330 and 360 pc). The CMD reveals the incredible precision of the OmegaCAM photometry and Gaia parallaxes and very well defined sequences, indicative of very young populations (in comparison to the background of galactic stars). The H$\alpha$ photometry confirms that very few stars show H$\alpha$ excess, usually used as sign of on-going accretion in pre-Main Sequence stars~\citep[e.g.][]{ka15}. This is in agreement with the fact that stars in the Vela OB2 region are older then 10~Myr and, hence, no accretion discs are expected~\citep[][]{fe10}.
The age for each cluster, obtained using solar metallicity models~\citep[][]{spina14} from \citet{Bressan2012} and $A_V=0.18$~\citep[in agreement with][]{sacco15}, are listed in Tab.~\ref{tab:properties}. 
The fit of the CMD for each cluster is shown in Fig.~\ref{fig:parcmdha}.

\section{Discussion} \label{Sec:Conc}
\citet[][]{jeff14} find 2 kinematically distinguished populations around  $\gamma^2$ Velorum, an old (10~Myr), bound ($\sigma_A=0.34\pm0.16$~kms$^{-1}$) population A
with a mean RV$_A=16.73\pm0.09$~kms$^{-1}$ and centrally concentrated around $\gamma^2$ Velorum. Population B is more dispersed, slightly younger and has a broader dispersion of $1.60\pm0.37$~kms$^{-1}$ and is offset in RV from the first by 2~kms$^{-1}$.
\citet[][]{jeff14} propose several scenarios to explain the presence of two populations, concluding that population A is the remnant of a bound cluster 
formed around the massive binary, and population B is a dispersed population from the wider Vela OB2 association. Later,~\citet[][]{sacco15} in a GES-based RV study of the
stellar  population in the cluster NGC 2547, discovered 15 stars that appear kinematically similar to population B of Gamma Vel and are located 2 deg south of $\gamma^2$ Velorum.

Thanks to the combination of GAIA DR2, OmegaCAM photometry and the GES measurements, we are able now to investigate with greater precision the stellar populations in the area.
By combining parallaxes, PMs and radial distribution of the stars in the area, we identify 6 different clusters.
According to our study, NGC 2547 (Cl3) is located at a distance of 393~pc~\citep[hence larger  than the distance of 340pc adopted by][]{sacco15}, shows an age of 30Myr, $\mu_\alpha$=$-8.57\pm0.29$mas/y
and $\mu_\delta$=$4.27\pm0.27$mas/y and RV=$12.29\pm3.71$kms$^{-1}$ in agreement with the population of NGC 2547 A from~\citet[][]{sacco15}. The main cluster around $\gamma^2$ Velorum,
~\citep[Gamma Vel A/population A in][]{jeff14} corresponds to our Cl4. It shows a distance of 351~pc, age 
of 10 Myr~ \citep[older than the estimated age of the Wolf-Rayet star $\approx 5.5\pm1$ Myr;][]{eld09}, $\mu_\alpha$=$-6.40\pm0.20$mas/y and $\mu_\delta$=$9.56\pm0.32$mas/y and RV=$18.50\pm4.50$kms$^{-1}$.
Hence, the main population of the Cl3 (NGC2547 A) and Cl4 (Gamma Vel A) clusters are separated by $\approx 44$~pc in 3-D. We emphasise here that thanks to the parallaxes and
PMs information available with GAIA DR2, we could naturally remove any contamination from a population ``B'' in the RV distribution of both the clusters NGC 2547 and Gamma Vel.
Noticeably,  Cl2 (clearly identified both in PM and RA/$\delta$ space) has properties in term of PMs and age similar to Cl4 (Gamma Vel A), but it is located at a larger distance (413~pc). 
RV measurements from GES are available for some of the stars belonging to Cl2 and selected in the PM space. 
These stars show an average RV=$20.36\pm0.6$kms$^{-1}$, which is in good agreement with NGC 2547B studied by~\citep[][]{sacco15}.
Cl2 has a well-defined centre in the $\alpha$ and $\delta$ space and is located at a 3-D distance of 64~pc from Gamma Vel and 24~pc from NGC 2547. 
It is likely that such stars were identified as NGC 2547 B by~\citet[][]{sacco15}. Our data seems to indicate that they do not belong to Gamma Vel even
if they show the same age as they are located at a much larger distance.

Cluster 6 (violet) is very interesting as it appears predominantly in the PM space, but is much looser in the apparent sky position. It is, however, very well constraint in parallax 
and located at a distance of $\approx 330$~pc.  Most interesting is that it has an age, RV and PM properties which are similar to that of Cl 3, i.e. NGC 2547, although its centre is located 65~pc
away from this cluster in 3-D. 

Hence, to summarise, Cl2 and Cl4 (Gamma Vel) show same age and PM properties but a difference of RV of $\approx 6$~kms$^{-1}$ and a 3-D distance of 64~pc from each other. Similarly
Cl6 and Cl3 (NGC 2547) do show same age and similar PM and RV ($\approx 12$~kms$^{-1}$) but are located at a 3-D distance of 65~pc from each other.
Cl1 also show PM and age similar to Cl4 (Gamma Vel) but is located at a 3-D distance of 52~pc from it and 31~pc from the much older Cl3 (NGC 2547).
Interestingly, it is located at a 3D distance of 28~pc from the coeval Cl2 which show similar position in the PM space. No RV information are available for Cl1.

Overall, our study put the stellar populations in the Vela OB2 complex under a new prospective. Thanks to the unique capabilities offered by the Gaia-Dr2 catalogues,
it is indeed proved that the stellar ``B'' populations originally discovered in Gamma Vel and NGC 2547 by~\citet[][]{jeff14} and \citet[][]{sacco15} respectively, are in
fact stars belonging to a complex set of clusters overlapping in space over an area of almost 100~pc. The RV information available for each cluster, combined with the
PMs, indicate that all the clusters move on an almost parallel trajectory from South-West to North-East. Based on the available data and tracing back the trajectory of
the stars for which RV and PMs are available, we could not find a common location from where the clusters could have originated 10 to 30 Myr ago. 

It is interesting to note that OB associations show spatial and kinematic substructure in the form of groups 
and clumps that potentially will survive as the entire association disperses. In this context, the initial 
conditions of the complex of clusters found in this work are likely to be very similar to that seen 
(structurally and kinematically) in younger OB associations such as Cygnus OB2~\citep[][]{w16} and Scorpius-
Centaurus~\citep[][]{w18}. It is hence reasonable to hypothesize that such regions might resemble the progenitor of the distribution of Vela OB2 co-moving clusters at 5-10 Myr time. In more distant star forming regions, particularly outside the Milky Way, without accurate PM information clusters of this type would be hard to distinguish from one another. This could explain the spread of ages observed e.g. in young star forming regions in the Magellanic Clouds~\citep[][]{de17}.
While we can not discard that the clusters are in fact sub-clusters formed inside the same molecular clouds in a 
fragmented and filamentary structure, more
detailed kinematic information will be needed to re-construct the star formation
history of this complex region in order to firmly confirm if, where and when the stars in these clusters formed.  


\section*{Acknowledgements}

This research has made use of the services of the ESO Science Archive Facility.
This work has made use of data from the European Space Agency (ESA) mission
{\it Gaia} (\url{https://www.cosmos.esa.int/gaia}), processed by the {\it Gaia}
Data Processing and Analysis Consortium (DPAC,
\url{https://www.cosmos.esa.int/web/gaia/dpac/consortium}). Funding for the DPAC
has been provided by national institutions, in particular the institutions
participating in the {\it Gaia} Multilateral Agreement. Based on data products from observations made with European
Southern Observatory Telescopes at the La Silla Paranal Observatory under programme ID 188.B-3002. These data products have
been processed by the Cambridge Astronomy Survey Unit (CASU)
at the Institute of Astronomy, University of Cambridge, and by the
FLAMES/UVES reduction team at INAF/Osservatorio Astrofisico
di Arcetri.




\bibliographystyle{mnras}
\bibliography{library} 








\bsp	
\label{lastpage}
\end{document}